\documentclass[sigconf]{acmart}
\usepackage{subcaption}
\usepackage{multirow}
\usepackage{xcolor}
\definecolor{mygreen}{rgb}{0, 0.45, 0}

\AtBeginDocument{%
  }

\copyrightyear{2026}
\acmYear{2026}
\setcopyright{cc}
\setcctype{by}
\acmConference[ETRA '26]{2026 Symposium on Eye Tracking Research and Applications}{June 01--04, 2026}{Marrakesh, Morocco}
\acmBooktitle{2026 Symposium on Eye Tracking Research and Applications (ETRA '26), June 01--04, 2026, Marrakesh, Morocco}
\acmDOI{10.1145/3797246.3803043}
\acmISBN{979-8-4007-2519-7/2026/06}

\begin{document}

\title[Gaze Prediction: Quantifying Performance Variability and Extreme-Case Errors]{Gaze Prediction as Time-Series Forecasting for Virtual Reality Applications: Quantifying Performance Variability and Extreme-Case Errors}

%% full authors here 
\author{Kateryna Melnyk}
\email{k_m825@txstate.edu}
\orcid{0000-0002-0244-9878}
\affiliation{%
  \institution{Texas State University}
  \streetaddress{601 University Dr, San Marcos}
  \city{San Marcos}
  \state{Texas}
  \country{USA}
}

\author{Lee Friedman}
\email{l_f96@txstate.edu}
\orcid{0000-0002-6385-1035}
\affiliation{%
  \institution{Texas State University}
  \streetaddress{601 University Dr, San Marcos}
  \city{San Marcos}
  \state{Texas}
  \country{USA}
}

\author{Oleg Komogortsev}
\email{ok@txstate.edu}
\orcid{0000-0001-7890-8842}
\affiliation{%
  \institution{Texas State University}
  \streetaddress{601 University Dr, San Marcos}
  \city{San Marcos}
  \state{Texas}
  \country{USA}
}

\renewcommand{\shortauthors}{Melnyk et al.}

\begin{abstract}
Gaze prediction is essential for addressing motion-to-photon latency and ensuring seamless foveated rendering in Virtual Reality. The reliability of gaze forecasting is highly sensitive to individual differences and the eye movements being predicted. We evaluate recurrent, transformer-based, and classification-guided architectures to assess their generalization capabilities across oculomotor events. Using the GazeBase VR and Meta Quest Pro datasets, we analyzed the relationship between the median ($P_{50}$) and high-percentile ($P_{95}$) error profiles across subjects. The analysis reveals significant performance variability, showing that subjects with low $P_{50}$ errors do not always exhibit the lowest extreme-case errors. Consequently, low median errors do not guarantee the robustness of the utilized solution. We discuss inference performance and address the class imbalance problem in short-term gaze prediction. These results identify a gap in standardized evaluation methods, necessitating a shift toward $P_{95}$-focused, subject-specific metrics to develop reliable and perceptually stable gaze-contingent systems.
\end{abstract}

%% The code below is generated by http://dl.acm.org/ccs.cfm.

%% The use of keywords is deprecated; do not include them in your source material.
%% \keywords{Eye Movement Prediction, Evaluation Metrics, Time-Series Forecasting}

\maketitle
\section{Introduction}
\label{sec:introduction}
Eye tracking (ET) is a sensor technology with a wide range of applications \cite{punde2017app} spanning from improving the user experience in virtual/augmented reality (VR/AR) \cite{adhanom2023eye} to enhancing driver assistance systems \cite{khan2019gaze} and assistive technologies for individuals with physical or cognitive disabilities \cite{paing2022design, kotseruba2024understanding}. Although advantageous, ET integration introduces real-time delays that demand efficient resource usage, particularly for latency-sensitive workloads such as foveated rendering (FR) and eye-controlled interfaces. Delays of even a few milliseconds (ms) can be perceptually significant, often leading to visual discomfort \cite{stauffert2020latency}. Gaze prediction serves to compensate for this latency by anticipating future eye positions.

The design of a gaze prediction solution is influenced by the prediction interval (PI), the time between the current and future positions. The task is classified as a short- or long-term prediction ($>100$ ms) based on the PI, and as a computer vision or time series forecasting task (TSF) based on the input data modalities. Short-term gaze prediction optimizes FR, a technique that renders the foveal region at full resolution while reducing the level of detail in peripheral areas without sacrificing an immersive experience. Most modern head-mounted displays (HMDs) sample gaze at 90 Hz \cite{stein2021comparison}. To mitigate the cases where the system cannot produce a gaze sample within the 11.1 ms window, forecasting future positions helps to maintain smooth rendering \cite{albert2017latency}. 

Evaluation of subject-dependent systems in real-time is an analytically complex task. Although the median error ($P_{50}$) is a stable measure of average performance, the $P_{95}$ analysis is critical to quantify peak inaccuracies. High errors in the 95th percentile cause incorrect rendering, compromise immersion, and induce user confusion \cite{li2024less}. Framing the task as a short-term TSF problem, we explore the performance of Transformers and Recurrent Neural Networks (RNNs) in gaze prediction, as both are designed to capture sequential dependencies \cite{lim2021timeseries}. We used gaze-tracking signals to predict positions within a 44 ms PI across three models: a long short-term memory (LSTM) network, a contrastive transformer encoder forecaster (CTEF), and a classifier-predictor network (ClPr). Model performance was assessed in both typical and tail cases, accounting for eye movement types (fixations, saccades, and post-saccadic intervals) and subject variability to provide practical insights into model selection.
 
\section{Literature Review}
Gaze prediction relied on model-driven approaches before deep learning (DL) became the standard. LSTMs have since been a common data-driven approach, though recent research has increasingly turned its attention to Transformer-based models  \cite{vaswani2017attention}. This transition is motivated by the non-stationary nature of gaze behavior, influenced by task design or the user's psychophysiological state. For example,  \citet{pavel2025patchfusionvr} demonstrated that the PatchFusionVR multi-task model, which integrates heart rate and galvanic skin response signals, improves prediction accuracy by using these physiological dynamics. Although RNNs excel at sequence modeling, their recursive nature often struggles with the abrupt shifts inherent in gaze patterns. Transformers address these limitations via self-attention, which provides the robustness needed for non-linear ET signals \cite{zeng2023transformers}. Simpler models may still outperform them when data properties or task constraints match their specific strengths \cite{wen2023transformers}.

\citet{xu2018gaze} proposed a hybrid architecture to predict gaze in immersive $360^{\circ}$ videos. They extract features from saliency maps and images using a convolutional neural network (CNN) and employ an LSTM to encode gaze paths to predict gaze displacement over time. \citet{hu2019sgaze} proposed SGaze, an eye-head coordination model for real-time prediction. They demonstrated that the correlation between head rotations and gaze positions is strong enough to predict gaze using only head poses. This work was extended through DGaze \cite{hu2020dgaze}, a multi-modal CNN that combines object positions, head velocity, and saliency features along with an ET-augmented version, DGaze\_ET. The findings indicated that for PIs between 50 ms and 1 s, gaze data becomes ineffective for forecasting horizons longer than 1 s. Later, they proposed FixationNet \cite{hu2021fixationnet} for fixation prediction, showing that forecasting long-duration events (>400 ms) is more challenging than short-duration ones.

\citet{mazzeo2021deep} compared several gaze estimation and prediction architectures, finding that a baseline LSTM outperformed the more advanced Informer for the prediction task.  \citet{rolff2022gazetransformer} proposed GazeTransformer, which fuses gaze, head, task, and image data. With a 150 ms PI, they observed that including image data yielded no significant performance improvements. This suggests that inertial measurement unit (IMU) and eye movement data alone are sufficient to capture gaze dynamics, supporting the use of more computationally efficient model variants. \citet{illahi2022real} developed a lightweight two-branch LSTM for low-latency gaze prediction in cloud-based VR applications. The model processes 200 ms sequences of gaze coordinates and rotational kinematics signals. Using PIs from 22 to 150 ms, they found that gaze data alone is a strong predictor, but performance improves when more features are added, such as gaze velocity or HMD rotations. Their generalization analysis suggests that optimal input sequence lengths and prediction windows are dataset-dependent and require corresponding adjustments.

While ML-based approaches are prevalent, model-driven solutions remain a critical area of advancement, as they provide faster inference and the interpretability that data-driven models typically lack. \citet{arabadzhiyska2017saccade} developed a polynomial model to predict saccade landing positions based on direction and amplitude for gaze-contingent rendering. Their analysis demonstrated that, while a generalized model is functional, personalization significantly improves accuracy. By applying subject-specific parameter tuning to the Oculomotor Plant Mathematical Model (OPMM) \cite{katrychuk2022study}, \citet{katrychuk2025oculomotor} developed OPMM in Kalman Filter form (OPKF) optimized for personalized continuous gaze prediction. \citet{melnyk_gp} evaluated the OPKF against LSTM and transformer-based networks, indicating that DL models outperform it across various metrics. Although gaze prediction is advancing, a high-performing model does not guarantee an efficient Tracked Foveated Rendering (TFR) system. A seamless user experience depends on optimizing the entire end-to-end pipeline. To address this, Liu et al. developed FovealNet \cite{liu2025fovealnet}, an AI-based gaze tracking framework. While traditional models prioritize average accuracy, FovealNet minimizes 'long-tail' errors through a performance-aware training strategy, yielding a 1.42x speedup and a $13\%$ increase in perceptual quality.

\section{Methodology}
\subsection{Datasets}
\subsubsection{Meta Quest Pro VR Dataset.} 
ET data were captured at 90 Hz from 78 students at Texas State University (mean age 19.5, range 18 to 28) using a Meta Quest Pro VR headset. The dataset and experimental design are described in detail by \citet{aziz2024evaluation}. We used two random saccade (RAN) tasks in which the target was randomly moved along a grid that spanned $\pm25^{\circ}$ horizontally and $\pm20^{\circ}$ vertically. RAN 127 and RAN 63 are identical in design, but vary in background luminance: light (127/255) and dark gray (63/255).

\subsubsection{GazeBase VR Dataset.} ET data were collected at 250 Hz from 407 participants (age range 18–58, predominantly 18–25) using a HTC Vive VR headset. The dataset and experimental design are described in detail by \citet{lohr2023gazebasevr}. We used a RAN task in which a target moved to random positions within a $\pm15^{\circ}$ horizontal and $\pm10^{\circ}$ vertical field. The task consists of 79 stimulus movements (80 fixation periods).

\subsection{Prediction Models}
\subsubsection{Model Architectures and Selection Rationale.} We selected these architectures to investigate distinct paths for error reduction: a transformer-based model incorporating contrastive learning, and a multitask learning (MTL) model employing explicit oculomotor event recognition. Our goal is to determine whether these specialized mechanisms can effectively mitigate the high-magnitude $P_{95}$ errors typical of non-stationary gaze data.

\subsubsection{Long Short-Term Memory Network - LSTM} As a baseline, we implemented a 2-layer bidirectional LSTM with a hidden dimension ($d_{\text{h}}$) of 128. To enforce feature invariance to input noise, the model employs a dual-loss function: exponential weighted mean absolute error ($L_{1}$) for the forecasting task and normalized temperature-scaled cross-entropy (NT-Xent, $\tau=0.07$) to regularize the encoder.

\subsubsection{Contrastive Transformer Encoder Forecaster - CTEF} The CTEF is a lightweight Transformer implementation that integrates a self-supervised contrastive learning branch to better penalize prediction inaccuracies during high-velocity ocular events. The architecture employs a streamlined, 4-layer encoder-only design ($d_{\text{model}}=128, n_{\text{head}}=4, d_{\text{ffn}}=512$) that maps raw features to a higher-dimensional embedding space. It utilizes sinusoidal positional encoding and multi-head self-attention to capture temporal dependencies before a two-stage MLP (multi-layer perceptron) forecasting head flattens and projects the features to the target output length. To distinguish gaze patterns from jitter, we implement a self-supervised task in which a ``positive'' pair is created via Gaussian noise augmentation ($\sigma=0.05$). The model is optimized via $L_{\text{total}} = L_{\text{MAE}} + \lambda L_{\text{NT\_Xent}}$ ($\lambda=0.5$), forcing the model to learn noise-invariant representations.

\subsubsection{Classifier-Predictor Network - ClPr} The ClPr treats gaze prediction as a MTL problem. It uses a shared hybrid encoder consisting of a 4-layer Temporal Convolutional Network (kernel size 7) for local feature extraction and a 3-layer LSTM ($h_{\text{d}}$ = 128) for long-range dependencies. It branches into a regression head for forecasting and a classification head for event recognition. To focus learning on physiologically meaningful data, a valid-class mask is applied to cross-entropy loss ($L_{\text{CE}}$), forcing the model to ignore artifacts and prioritize fixations and saccades. The model is optimized via: $L_{\text{total}} = w_{\text{pred}} L_{\text{MSE}} + w_{\text{class}} L_{\text{CE}}$. Extended model details are provided in the Appendix.

\subsection{Data Pre-processing}
We classified ET signals into eye movement events using the FKM algorithm (detailed in \cite{melnyk_gp}), an approach based on the MNH method \cite{mnh}. The MNH classifies ET signals by first applying an adaptive threshold to identify artifact blocks, followed by dual-threshold radial velocity detection to categorize the valid data into saccades and fixations. The FKM incorporates refined logic to more accurately identify artifacts, a routine to correctly classify ``overlapping'' saccades as a single saccade event, and a reclassification of microsaccades ($<1^{\circ}$) that occur within the fixation as part of the fixation. We adapted the FKM algorithm for the Meta Quest Pro (90 Hz) and GazeBase VR (250 Hz) datasets, as it was originally developed for 1000 Hz signals. All preprocessing steps, such as filtering, interpolation, and blink detection, are summarized along with their parameter settings in the Supplemental Materials.

We performed a data ablation study to identify the most effective input modalities among gaze positions, velocities, accelerations, jerks, heading angles, and magnitudes of velocity and acceleration. We found that forecasting changes in positions $\Delta x_{\text{t}+\text{PI}} = x_{\text{t}+\text{PI}} - x_{\text{t}}$ was more effective than predicting absolute gaze positions. We treated the size of the input and output window as hyperparameters, testing the lengths of the input window between 6 and 20 samples (66 and 220 ms). Regarding output lengths, forecasting a sequence from $x_{\text{t}-\text{k}+\text{PI}}$ to $x_{\text{t}+\text{PI}}$, where $k\in[1, PI]$, outperformed predicting only the final step $x_{\text{t}+\text{PI}}$. Based on these results, we define the final input configurations for each model: LSTM used [positions, velocities, heading angles] (window size $ws=12$); CTEF used [positions, velocities] ($ws=6$); and ClPr used [velocities, heading angles, classification labels] ($ws=6$). All models predict the sequence of the next PI steps.

\subsection{Training}
To evaluate model generalization and quantify performance variability, we used a 5-fold subject-independent cross-validation. For each fold, the datasets were partitioned using a subject-wise random split to ensure that data from the test subjects were not present during the training phase. The training/testing splits per fold were 56/12 subjects (112/24 recordings) for Meta Quest Pro and 210/37 subjects for GazeBase VR. To prepare the data, we employ an adaptive windowing strategy that adjusts the overlap step at signal segments containing saccade samples. Traditional fixed-step windowing, common for 1000 Hz data \cite{melnyk_gp, aziz2023practical}, was not suitable for lower sampling rates. It results in an over-representation of fixation-dominated windows at these rates. This adaptive approach addresses the class-imbalance problem by oversampling the minority classes (specifically saccades and pre- and post-saccadic periods) to enable more robust feature learning. Windows were shuffled before partitioning into batches of 32. We used the PyTorch Lightning framework \cite{falcon2019pytorch} to facilitate a more efficient and stable training. The Adam optimizer was used with a learning rate of 0.0003. The 44 ms PI was chosen to offset the 30--50 ms end-to-end latency typical of modern HMDs. Visualizations are provided in the Appendix, with additional plots for 22 and 66 ms PIs available in the Supplementary Materials. All experimental data, source code, and supplemental materials are available at: https://hdl.handle.net/10877/23835.

\subsection{Evaluation}
In this study, we examine the average performance and $P_{95}$ error distributions, reporting all errors in degrees of visual angle ($^{\circ}$). To characterize overall population performance, we pooled out-of-sample test predictions from all five folds, resulting in a total evaluation pool of 60 unique subjects for Meta Quest Pro \cite{aziz2024evaluation} and 185 for GazeBase VR \cite{lohr2023gazebasevr} datasets. Analyzing $P_{95}$ highlights error tail performance, which is critical for identifying models that remain robust under stress conditions or when encountering unusual outliers. For our summary tables, we report the median and $P_{95}$ per-sample errors, along with the grand median values for the per-subject $P_{50}$ and $P_{95}$ errors. This combination provides a robust estimate of both average performance and individual variability. Errors were analyzed for different eye movement events: fixations, large-amplitude saccades ($>2^{\circ}$), and small-amplitude saccades, including critical evaluation periods (CEPs) \cite{aziz2023practical}, defined as 110 ms post-saccadic intervals. The Appendix includes additional metrics such as error cumulative distribution functions (cdfs) and per-subject $P_{50}$ and $P_{95}$	error profiles to provide a more comprehensive evaluation.

\section{Results}
\subsection{Gaze Prediction as a Function of Eye Movement Type}
Table~\ref{tab:comparison_QP_GBVR} compares model performance across two datasets. The PI values align with the sampling rates: 44 ms (4 samples at 90 Hz, Quest Pro) and 40 ms (10 samples at 250 Hz, GazeBase). LSTM achieved the lowest error for small saccades. CTEF outperformed all other models in the remaining categories. As documented in the Appendix, prediction errors expectedly increased with larger PI values. The Appendix provides further detail through performance visualizations of error distributions for eye movement events, including cdfs and boxplots.

\begin{table*}[h]
    \centering
    \caption{Comparison of Per-Sample Prediction Errors Across Event Types for Quest Pro and GazeBase VR Datasets. Bold values (highlighted in green) indicate the lowest prediction error for each category.}

    \begin{tabular}{|c|l|l|c|c|c|c|c|c|c|c|}
        \hline
        \multirow{2}{*}{\textbf{Dataset}} & \multirow{2}{*}{\textbf{PI}} & \multirow{2}{*}{\textbf{Model}} & \multicolumn{2}{|c|}{\textbf{Fixation}} & \multicolumn{2}{c|}{\textbf{Large Saccade}} & \multicolumn{2}{c|}{\textbf{Small Saccade}} & \multicolumn{2}{c|}{\textbf{CEP}} \\
        \cline{4-11}
        & & & $P_{50}$ & $P_{95}$ & $P_{50}$ & $P_{95}$ & $P_{50}$ & $P_{95}$ & $P_{50}$ & $P_{95}$ \\
        \hline

        \multirow{3}{*}{Quest Pro} &
        \multirow{3}{*}{44} & LSTM & 0.47 & 1.64 & 3.9 & 10.02 & \textbf{\textcolor{mygreen}{1.13}} & \textbf{\textcolor{mygreen}{1.828}} & 0.68 & 3.32 \\
        \cline{3-11} 
        & & \textbf{\textcolor{mygreen}{CTEF}} & \textbf{\textcolor{mygreen}{0.46}} & \textbf{\textcolor{mygreen}{1.55}} & \textbf{\textcolor{mygreen}{3.88}} & 10.14 & 1.15 & 1.83 & \textbf{\textcolor{mygreen}{0.67}} & \textbf{\textcolor{mygreen}{3.18}} \\
        \cline{3-11}
        & & ClPr & 0.55 & 1.81 & 4 & \textbf{\textcolor{mygreen}{9.97}} & 1.24 & 1.98 & 0.76 & 3.61 \\
        \hline

        \multirow{3}{*}{GazeBase VR} &
        \multirow{3}{*}{40} & LSTM & 0.179 & 0.65 & 3.8 & 8.99 & \textbf{\textcolor{mygreen}{0.8}} & 2.09 & 0.252 & 1.57 \\
        \cline{3-11} 
        & & \textbf{\textcolor{mygreen}{CTEF}} & \textbf{\textcolor{mygreen}{0.174}} & \textbf{\textcolor{mygreen}{0.59}} & \textbf{\textcolor{mygreen}{2.7}} & \textbf{\textcolor{mygreen}{8.67}} & 0.85 & \textbf{\textcolor{mygreen}{2.06}} & \textbf{\textcolor{mygreen}{0.251}} & 1.46 \\
        \cline{3-11}
        & & ClPr & 0.24 & 0.84 & 3.6 & 8.84 & 0.86 & 2.15 & 0.32 & \textbf{\textcolor{mygreen}{1.44}} \\
        \hline 
    \end{tabular}
    \label{tab:comparison_QP_GBVR}
\end{table*}

\subsection{Gaze Prediction as a Function of Individual Differences}
A Weibull-based distribution fitting approach was used to calculate $P_{50}$ and $P_{95}$ for each subject, providing a robust estimation of both median performance and tail-end error characteristics. Table \ref{tab:comparison_QP_GBVR_p50_p95} compares these results by reporting the grand median (M) across the error distributions at the subject-level. We used Wilcoxon signed-rank tests to evaluate the significance of performance differences between models. A Bonferroni-corrected significance threshold of $\alpha=0.0167$ was used to account for the three pairwise comparisons made within each eye movement event.

Error rates are consistently lower on GazeBase VR compared to Quest Pro. For example, the $M(P_{50})$ per-subject fixation error for CTEF is approximately $64\%$ lower on the GazeBase VR than on the Quest Pro dataset. Error rates remain low during fixations. Large saccades continue to be the primary challenge, with $M(P_{95})$ errors exceeding $14^{\circ}$ on the Quest Pro dataset. CEP provides a representative measure that balances fixation stability against the unpredictability of saccades. The shift from $M(P_{50})$ to $M(P_{95})$ captures the performance degradation at the edge of the error distribution. For example, the Quest Pro median CEP error of approximately $\approx0.6^{\circ}$ climbs to over $4.0^{\circ}$ at the 95th percentile.

The metrics reported in Table \ref{tab:comparison_QP_GBVR} and Table \ref{tab:comparison_QP_GBVR_p50_p95} represent different levels of granularity. The error values in Table \ref{tab:comparison_QP_GBVR} reflect fine-grained, sample-wise fluctuations across the entire dataset. The per-subject metrics account for this variance by treating each subject as an independent observation unit. This approach ensures that the results reflect the generalized performance of the model across the user population.

\begin{table*}[h]
    \centering
    \caption{ Comparison of Per-Subject Errors Across Eye-Movement Events on the Quest Pro and GazeBase VR Datasets. M($P_{50}$) and M($P_{95}$) represent the grand median across all subjects. An asterisk $*$ denotes that the best-performing model achieved a statistically significant improvement ($p<0.0167$) against all other compared methods.}
    \begin{tabular}{|c|l|l|c|c|c|c|c|c|c|c|}
        \hline
        \multirow{2}{*}{\textbf{Dataset}} & \multirow{2}{*}{\textbf{PI}} & \multirow{2}{*}{\textbf{Model}} & \multicolumn{2}{|c|}{\textbf{Fixation}} & \multicolumn{2}{c|}{\textbf{Large Saccade}} & \multicolumn{2}{c|}{\textbf{Small Saccade}} & \multicolumn{2}{c|}{\textbf{CEP}} \\
        \cline{4-11}
        & & & $M(P_{50})$ & $M(P_{95})$ & $M(P_{50})$ & $M(P_{95})$ & $M(P_{50})$ & $M(P_{95})$ & $M(P_{50})$ & $M(P_{95})$ \\
        \hline

        \multirow{3}{*}{Quest Pro} &
        \multirow{3}{*}{44} & LSTM & 0.38 & 1.05 & 3.78 & \textbf{\textcolor{mygreen}{14.59}} & \textbf{\textcolor{mygreen}{1.16*}} & \textbf{\textcolor{mygreen}{2.29}} & \textbf{\textcolor{mygreen}{0.627}} & 4.02 \\
        \cline{3-11} 
        & & \textbf{\textcolor{mygreen}{CTEF}} & \textbf{\textcolor{mygreen}{0.37*}} & \textbf{\textcolor{mygreen}{1.01*}} & \textbf{\textcolor{mygreen}{3.73}} & 14.85 & 1.18 & 2.34 & 0.631 & \textbf{\textcolor{mygreen}{3.8*}} \\
        \cline{3-11}
        & & ClPr & 0.47 & 1.29 & 3.88 & 14.61 & 1.22 & 2.54 & 0.71 & 4.26 \\
        \hline

        \multirow{3}{*}{GazeBase VR} &
        \multirow{3}{*}{40} & LSTM & 0.14 & \textbf{\textcolor{mygreen}{0.498*}} & 2.73 & 12.6 & \textbf{\textcolor{mygreen}{0.71*}} & 1.94 & \textbf{\textcolor{mygreen}{0.377*}} & 3.08 \\
        \cline{3-11} 
        & & \textbf{\textcolor{mygreen}{CTEF}} & \textbf{\textcolor{mygreen}{0.13*}} & 0.501 & \textbf{\textcolor{mygreen}{2.27*}} & \textbf{\textcolor{mygreen}{11.24*}} & 0.72 & \textbf{\textcolor{mygreen}{1.94}} & 0.384 & 2.67 \\
        \cline{3-11}
        & & ClPr & 0.19 & 0.66 & 2.7 & 12.32 & 0.74 & 2 & 0.42 & \textbf{\textcolor{mygreen}{2.65*}} \\
        \hline
    \end{tabular}
    \label{tab:comparison_QP_GBVR_p50_p95}
\end{table*}

\subsection{Computational Complexity Analysis}
Although the models are not yet optimized for real-time execution, their relative computational complexity is compared using Multiply-Accumulate Operations (MACs) and total parameter counts (in millions). Specifically: (1) the LSTM (input size: $5\times12$) requires 27.09M MACs and 4.02M parameters; (2) CTEF (input size: $4\times6$) requires 6.05M MACs and 2.02M parameters; and (3) ClPr (input size: $4\times6$) requires only 4.49M MACs and 0.75M parameters. 

ClPr is the most computationally efficient of the three models based on the parameter count, whereas the LSTM and CTEF architectures require significantly higher MACs and memory. For comparison, the OPKF (from the study by \citet{melnyk_gp}) features only 13 parameters and substantially fewer MACs, making it the most suitable for strict real-time deployment. Reducing architectural complexity or window length will improve efficiency, but such changes require a careful trade-off with predictive accuracy.

\section{Discussion}
We demonstrate that average accuracy is an incomplete measure of gaze prediction performance, as $P_{95}$ error analysis is required to reveal critical insights into the model's worst-case behavior. The results indicate that model selection for short-term gaze prediction is a multi-faceted task requiring tuning across eye movement types. We chose these approaches under the theoretical assumption that their learning mechanisms handle temporal dependencies in ET data differently. Integration of adaptive windowing mitigated class imbalance across eye-movements. Through a contrastive learning branch and class-guided prediction, we improved the models' ability to distinguish saccades from hardware-induced artifacts and noise. CTEF emerges as the most robust architecture for minimizing edge-case errors among the three proposed. As shown in Tables \ref{tab:comparison_QP_GBVR} and \ref{tab:comparison_QP_GBVR_p50_p95}, CTEF achieves the lowest per-sample errors and demonstrates statistically significant improvements in the median M($P_{50}$) and worst-case M($P_{95}$) per-subject errors across most categories.

Both tables report lower overall errors for the GazeBase VR dataset. There is a marked difference in median spatial precision: 0.03--0.04$^{\circ}$ for GazeBase VR versus $0.65^{\circ}$ for Quest Pro. The lower sampling frequency of this hardware results in significantly less temporal density, limiting the models' ability to distinguish saccades from occasional artifacts or noise. These combined hardware factors, such as higher jitter and lower temporal resolution, create a lower ``predictive ceiling'' for the Quest Pro dataset. As an example, a lightweight LSTM \cite{melnyk_gp} evaluated on the 1,000 Hz GazeBase dataset \cite{griffith2021gazebase} achieved a $P_{50}=0.3^{\circ}$ fixation error, compared to the $0.45^{\circ}$ we observed in this study.

To relate error analysis to human visual perception, we included post-saccadic periods (CEPs) as part of the evaluation. Saccadic suppression typically begins $\sim$50 ms before saccade onset and ends 50–100 ms after saccade offset \cite{bidder1997comparison}. The timing of vision recovery is a function of the saccade amplitude, with recovery occurring later after larger saccades. During this window, artifacts or delayed updates caused by FR can be highly noticeable and cause user discomfort. CTEF is the most accurate during CEPs, based on both the $P_{50}$ and $P_{95}$ metrics (see Table \ref{tab:comparison_QP_GBVR}).

Although subject variability in eye movements \cite{risko2012curious, castelhano2008stable} and signal noise \cite{raju2024signal} are well-documented, their impact on gaze prediction is critical yet understudied. Models must incorporate this individual variance into their performance evaluations to be truly deployable \cite{melnyk_gp}. Analysis of $M(P_{95})$ and per-subject trends reveals the robustness of the model in cases sensitive to individual differences. In the Appendix, a comparison of subject-level $P_{50}$ and $P_{95}$ trends shows consistency across fixations and large saccades. The patterns decouple for CEPs and small saccades, suggesting that the architectures may prioritize different underlying properties of eye movements for prediction. While further interpretability studies are needed to confirm this, a closer look at the model feature reliance during specific gaze events remains a promising direction for future research.

We provide a brief analysis of computational costs, but a complete latency assessment requires evaluating the entire pipeline and optimizing the models on target hardware. Our findings suggest two critical goals for future gaze prediction: (1) narrowing the performance gap between top-performing and high-error subjects, and (2) minimizing the per-subject $P_{50}$ and $P_{95}$ error variance. One promising direction involves pre-training a model on a full dataset followed by personalized fine-tuning on 5 minutes of recent ET data (see \cite{melnyk_gp, katrychuk2025oculomotor} for a further discussion).

\subsection{Limitations}
In this paper, models utilize only gaze-tracking signals and their physiological derivatives. Although head movement data or graphical content is often cited as performance boosters \cite{illahi2022real}, gaze-only models can outperform those that integrate image data \cite{rolff2022gazetransformer}. This is partly due to overlapping mutual information between the input streams. For example, the strong correlation between gaze and head movement in dynamic scenes \cite{hu2020dgaze} suggests that the use of both may result in redundant feature spaces  without improving accuracy \cite{illahi2022real}. Computational efficiency remains a concern, as additional modalities can introduce significant system overhead. Gaze-only solutions offer low latency $\sim2$ ms \cite{illahi2022real}. Integration of saliency maps \cite{hu2019sgaze} or graphical data \cite{mazzeo2021deep} increases inference time to between 4.5 and 47 ms, falling short of real-time requirements.

Another limitation pertains to the RAN task. As a controlled task independent of scene content or user interest, its design provides a higher and more consistent saccade count. This helps models better capture saccadic dynamics, such as velocity-amplitude relationships and acceleration profiles, making our results more relevant to scenarios dominated by saccades than to smooth pursuit. This selection mitigates class imbalance. Lower sampling rates limit available data for specific events, such as 20 ms saccades, which may be represented by a single sample at 90 Hz or missed entirely.

Free-viewing tasks lead to unpredictable gaze behavior, mixed events, and noise, making automatic eye movement classification difficult. The classification of the two VR datasets in this study was a non-trivial undertaking that posed its own set of difficulties. A key objective for future research will be to identify smooth pursuit within gaze recordings and evaluate prediction accuracy separately during these events, as this area remains unaddressed. Another direction for future research involves designing tasks that incorporate both distinct saccades and smooth pursuit targets. Diversifying model training and testing across various ET scenarios, such as reading, horizontal saccade tasks, and video viewing, will be essential for real-world deployment.
 
This paper evaluates three DL architectures, including a multi-task classification and forecasting model. Although effective, the classification step introduces latency that complicates real-time application. We observed distinct trade-offs between classical approaches, such as the Kalman Filter Framework, and data-driven solutions. A well-calibrated OPKF computes real-time predictions efficiently, but its parameter optimization is often a complex task. Recent reports \cite{melnyk_gp, katrychuk2025oculomotor} indicate that ML models outperform OPKF. Augmenting the OPKF with ML components is a promising strategy to improve its performance. Transformers effectively handle unexpected gaze shifts, whereas LSTMs remain more efficient for medium-length sequences and smaller datasets.

\subsection{Future Work}
Future research should investigate Transformer variants and hybrid models that combine the advantages of attention mechanism with the sequential processing of RNNs. Such architectures can effectively enhance prediction despite input noise and signal artifacts with the proper learning techniques. Investigating model personalization through subject-specific embeddings or meta-learning would allow the system to adapt to individual eye movement patterns. This study focused on evaluation rather than full real-time optimization. Achieving the performance necessary for real-world applications will require quantization and pruning, particularly for longer PIs and Transformers. Future work must also include full-pipeline inference metrics to account for preprocessing overhead, not reported here.

\section{Conclusion}
We have demonstrated that median performance metrics are insufficient to fully characterize model robustness under varying conditions. High $P_{95}$ errors represent critical failure points that significantly degrade the user experience for subjects with variable gaze patterns, as poor reliability is often masked by strong median results. Strong performance during fixations does not necessarily translate to success during post-saccadic intervals, highlighting a critical trade-off in model selection. Therefore, application-specific constraints and needs must guide the selection of gaze prediction algorithms for real-time use.

\section*{Ethics and Privacy Statement} Participants provided informed consent under a protocol approved by the Texas State University Institutional Research Board, acknowledging the sharing of de-identified data for research purposes. Both datasets contain no participant-identifying information. Although this research aims to improve algorithms that compensate for ET system latency, accurate gaze prediction poses ethical risks, such as potential content manipulation. The technical benefits of these models must be balanced against the risk of misusing them for coercive content delivery.

\begin{acks} 
Google Gemini (Flash 2.5) was used for grammar refinement and code syntax review. The study's conceptual design, methodology, and interpretations were developed exclusively by the authors, who are accountable for the final content.
\end{acks}

%% The next two lines define the bibliography style to be used, and the bibliography file.
\bibliographystyle{ACM-Reference-Format}
\bibliography{sample-base}

\setcounter{figure}{0}
\setcounter{section}{0}
\setcounter{table}{0}

\appendix
\label{sec:appendix}
\newpage
\textbf{Appendix}

\section{FKM}
The classification of eye movement events was performed using the FKM algorithm \cite{melnyk_gp}, which is based on the MNH algorithm \cite{mnh}. The transition from MNH to FKM involved significant refinements of the classification logic, specifically with regard to saccade detection. Although the FKM algorithm was primarily designed for 1000 Hz data, this study utilized two datasets: the GazeBase VR dataset with a sampling rate of 250 Hz \cite{aziz2024evaluation} and a Quest Pro dataset with a sampling rate of 90 Hz \cite{lohr2023gazebasevr}. To ensure temporal consistency, the inter-sample intervals (ISIs) were analyzed, and the eye-tracking (ET) signals were interpolated to maintain uniform spacing. The signals were processed using a zero-phase, fifth-order Butterworth low-pass filter with a 15 Hz cutoff before classification. The FKM algorithm categorizes signals into fixations, saccades, and artifacts. Due to the high frequency of blinks within the recordings, a blink detection procedure was implemented. Detailed visualizations and comprehensive reports on the FKM modifications are available in the Supplementary Materials.

\section{Prediction Models}
\subsection{Contrastive Transformer Encoder Forecaster - CTEF.} 
CTEF is designed to maintain high sensitivity to rapid eye movements using a similarity-matching task (contrastive learning). This allows the model to distinguish real eye movements from random sensor noise.

\begin{itemize}
    \item \textbf{Encoder and Forecasting Head:} 
    \begin{enumerate}
        \item \textit{Positional Encoding:} A fixed sinusoidal positional encoding captures relative temporal distances between gaze coordinates. This encoding is added to the projected input sequence before the dropout layer ($\text{dropout rate} = 0.1$).
        \item \textit{Encoder:} The architecture employs a 4-layer transformer encoder with an embedding dimension of $d_{\text{model}}=128$, 4 attention heads ($n_{\text{head}}=4$) and a feed-forward expansion factor of $d_{\text{ffn}}=512$ using ReLU activations.
        \item \textit{Forecasting Head:} A MLP that maps encoder features to future gaze coordinates via two stages:
        \begin{enumerate}
            \item \textit{Feature Compression}: The 2D encoder output ($d_{\text{model}} \times \text{seq\_len}_\text{x}$) is flattened and projected through a hidden layer of size $\frac{d_{\text{ffn}} \times \text{seq\_len}_\text{x}}{2}$ with ReLU activation.
            \item \textit{Output Projection:} A final linear layer maps the compressed representations to the target output dimensions ($c_{\text{out}} \times \text{seq\_len}_\text{y}$).
        \end{enumerate}
    \end{enumerate}
    \item \textbf{Contrastive Learning Strategy:} A self-supervised task is employed to enhance feature robustness against sensor noise:
    \begin{enumerate}
        \item \textit{Data Augmentation:} For every input sample, a positive pair is created by adding Gaussian noise ($\sigma = 0.05$). 
        \item \textit{Projection Head:} To prioritize global features over local noise, encoder outputs undergo global average pooling followed by a 2-layer MLP to produce a 128-dimensional embedding.
        \item \textit{Optimization:} The model minimizes a dual loss function:
        \[ L_{\text{total}} = L_{\text{MAE}} + \lambda L_{\text{NT-Xent}} \]
        where $\lambda = 0.5$ balances the accuracy of the forecast ($L_{\text{MAE}}$) with contrastive consistency ($L_{\text{NT-Xent}}$), using a temperature parameter $\tau = 0.07$ to scale similarity scores.
\end{enumerate}
\end{itemize}

\subsection{Classification-Predictor Network - ClPr}
The ClPr model treats gaze prediction as a multi-task learning problem, utilizing a shared hybrid encoder to perform simultaneously sequence forecasting and eye movement classification.

\begin{itemize}
    \item \textbf{TCN-LSTM Encoder:} A dual-stage feature extraction process is used to capture eye movement dynamics:
    \begin{enumerate}
        \item \textit{Temporal Convolutional Network (TCN):} A 4-layer TCN (kernel size 7, hidden dimension 128) with causal padding and ReLU activations extracts local features and detects rapid patterns, such as saccade onset.
        \item \textit{LSTM Module:} A 3-layer LSTM network with hidden size $h_{\text{d}}=128$ processes the TCN-refined features.
    \end{enumerate}
    \item \textbf{Task-Specific Heads:} The shared encoder output is processed by two specialized heads.
    \begin{enumerate}
        \item \textit{Regression Head:} A linear layer maps the final LSTM hidden state into the target output space ($\text{out}\_\text{dim} \times \text{seq\_len}_\text{y}$).
        
        \item \textit{Classification Head:} To identify eye movement types (e.g., fixations vs. saccades), the LSTM output is processed via adaptive average pooling and a linear layer predict categorical logits for eye movement classes.

        \item \textit{Selective Supervision (Masking):} To ensure the model learns from reliable physiological data, a valid-class mask is applied during training. The loss is calculated only for samples belonging to meaningful eye movement categories (label 1: fixations and label 2: saccades), effectively ignoring noisy events or artifacts (labels 3--5) identified by the FKM algorithm.
        
        \item \textit{Optimization:} The model is trained to minimize a weighted multi-task objective: 
        \[L_{\text{total}} = w_{\text{pred}} L_{\text{MSE}} + w_{\text{class}} L_{\text{CE}}\] 
        where $w_{\text{pred}} = 0.1$ and $w_{\text{class}} = 1.0$. 
    \end{enumerate}
\end{itemize}

\subsection{Adaptive Windowing}
To address class imbalance in lower-frequency ET recordings, we implemented a class-conditioned sliding window to oversample minority events. For the Quest Pro dataset, the algorithm uses a step size ($s_{\text{r}}$) of 5 during the fixation segments. Upon detection of saccades or post-saccadic periods, the step size is dynamically reduced to $s_{\text{r}}=1$. This prevents high-velocity ocular events, such as saccades and CEPs, from being underrepresented relative to frequent fixations.

\section{Visualizations: Gaze Prediction as a Function of Eye Movement Type}
\begin{figure*}[t]
    \centering

    % --- SUB-GROUP 1: Quest Pro ---
        \centering
        \begin{subfigure}{0.32\textwidth}
            \centering
            \includegraphics[width=\textwidth]{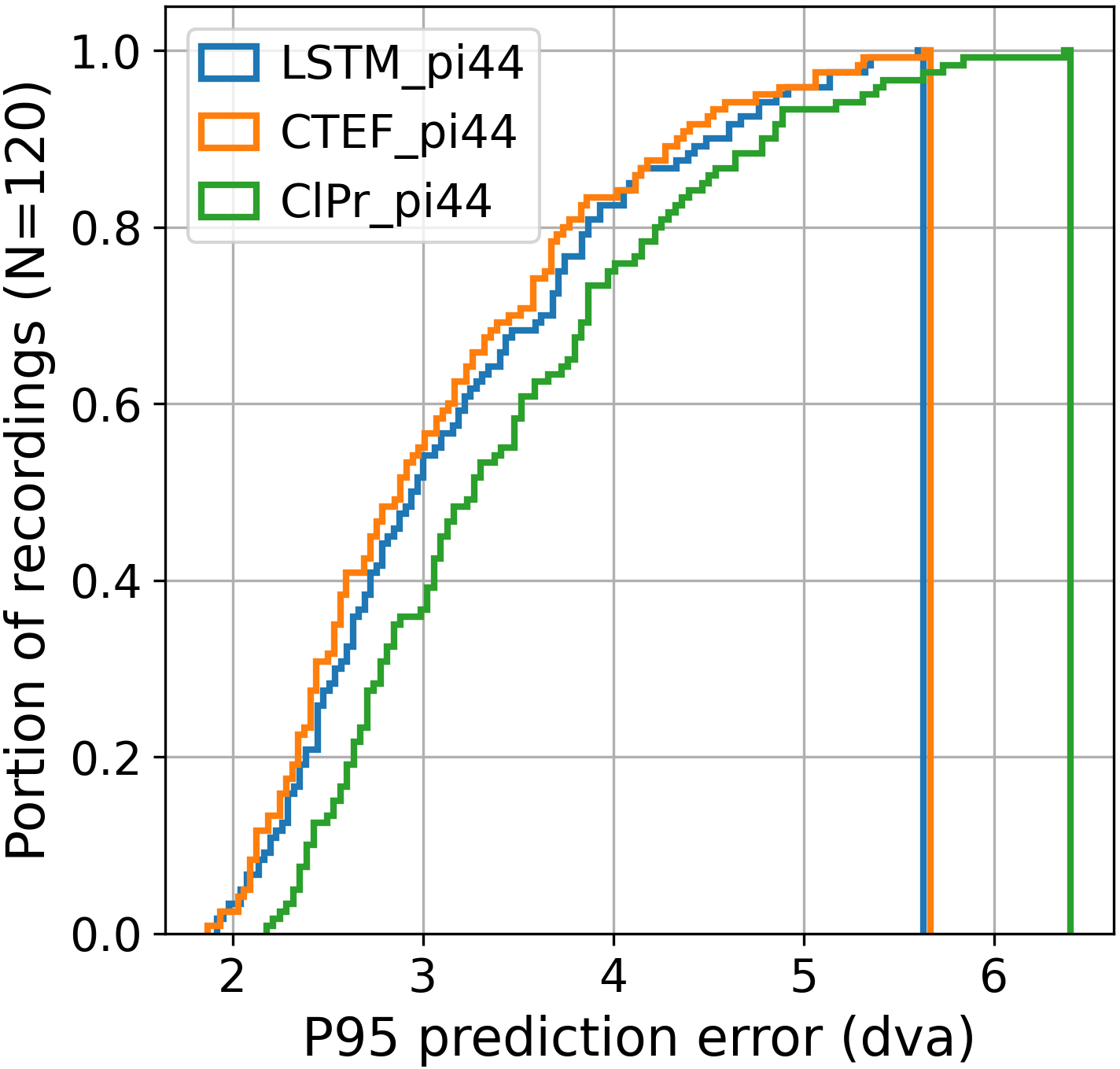}
            \captionsetup{justification=centering}
            \caption{\textbf{QP} \\ CDF of $P_{95}$ Event-Agnostic Errors}
            \label{fig:95_fullcdf_qp}
        \end{subfigure}
        \hfill
        \begin{subfigure}{0.32\textwidth}
            \centering
            \includegraphics[width=\textwidth]{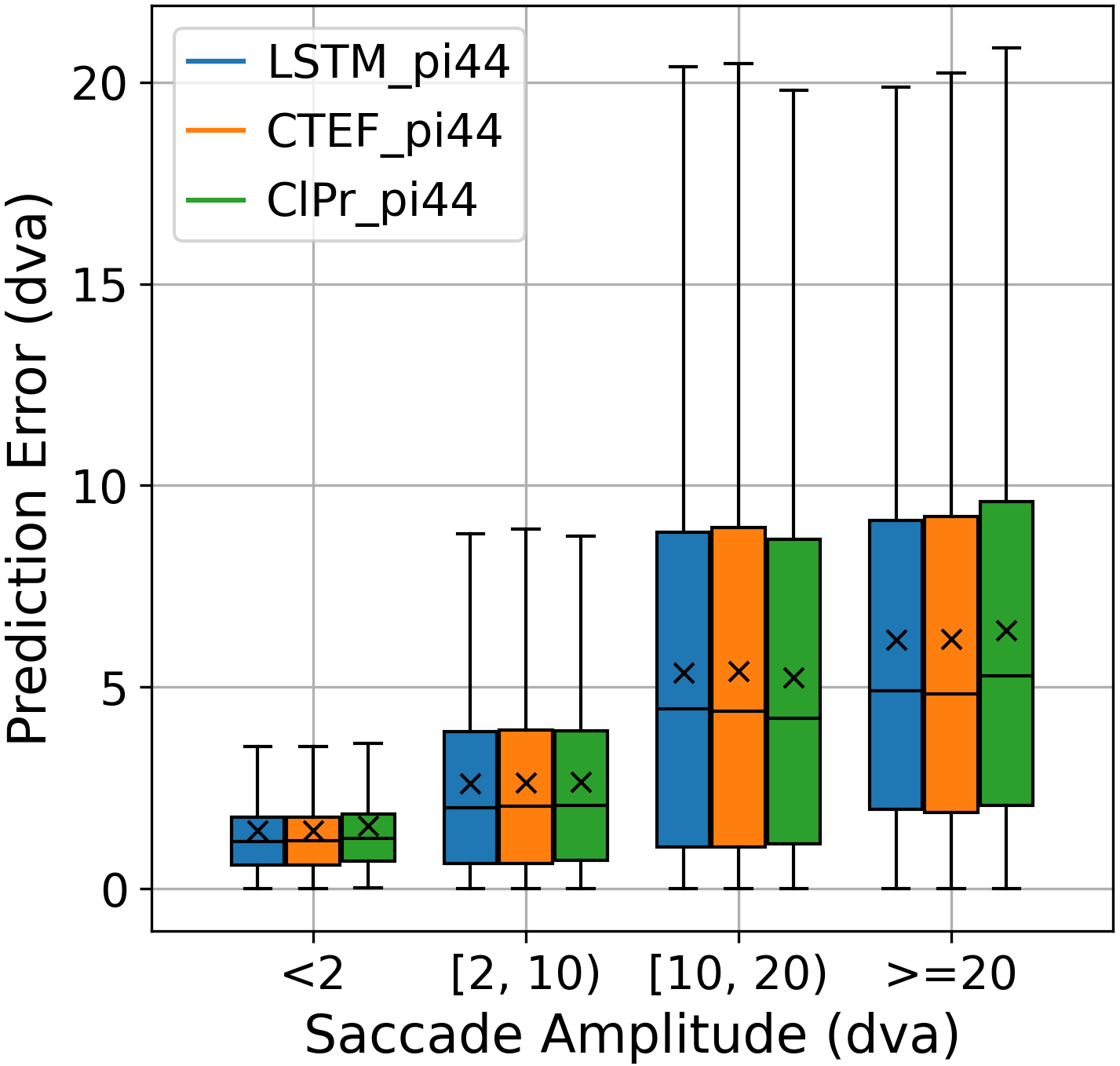}
            \captionsetup{justification=centering}
            \caption{\textbf{QP} \\ Saccade Prediction Errors}
            \label{fig:saccade_boxplot_qp}
        \end{subfigure}
        \hfill
        \begin{subfigure}{0.32\textwidth}
            \centering
            \includegraphics[width=\textwidth]{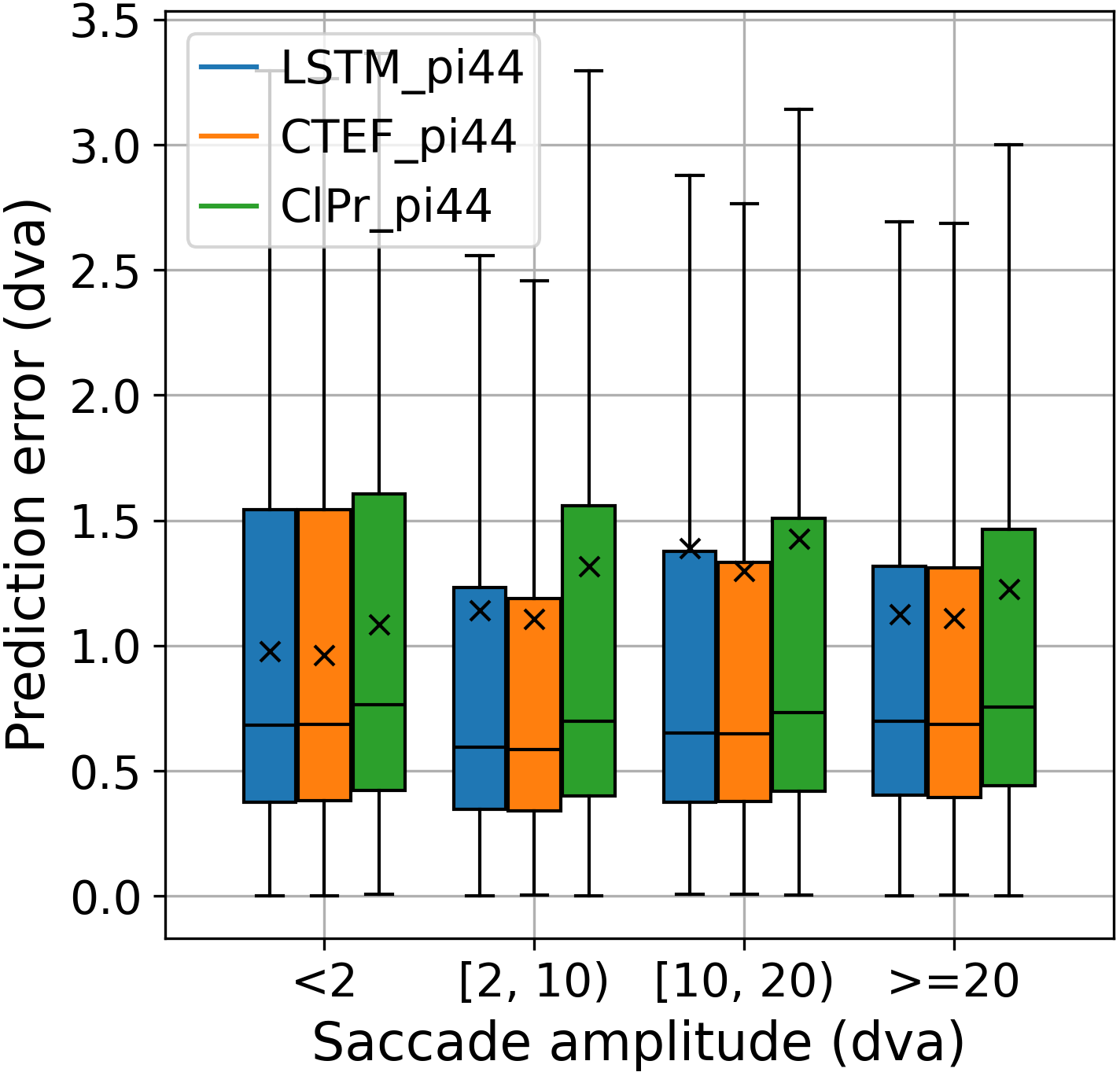}
            \captionsetup{justification=centering}
            \caption{\textbf{QP} \\ CEP Prediction Errors}
            \label{fig:cep_boxplot_qp}
        \end{subfigure}

    % --- SUB-GROUP 2: GazeBase VR ---
        \centering
        \begin{subfigure}{0.32\textwidth}
            \centering
            \includegraphics[width=\textwidth]{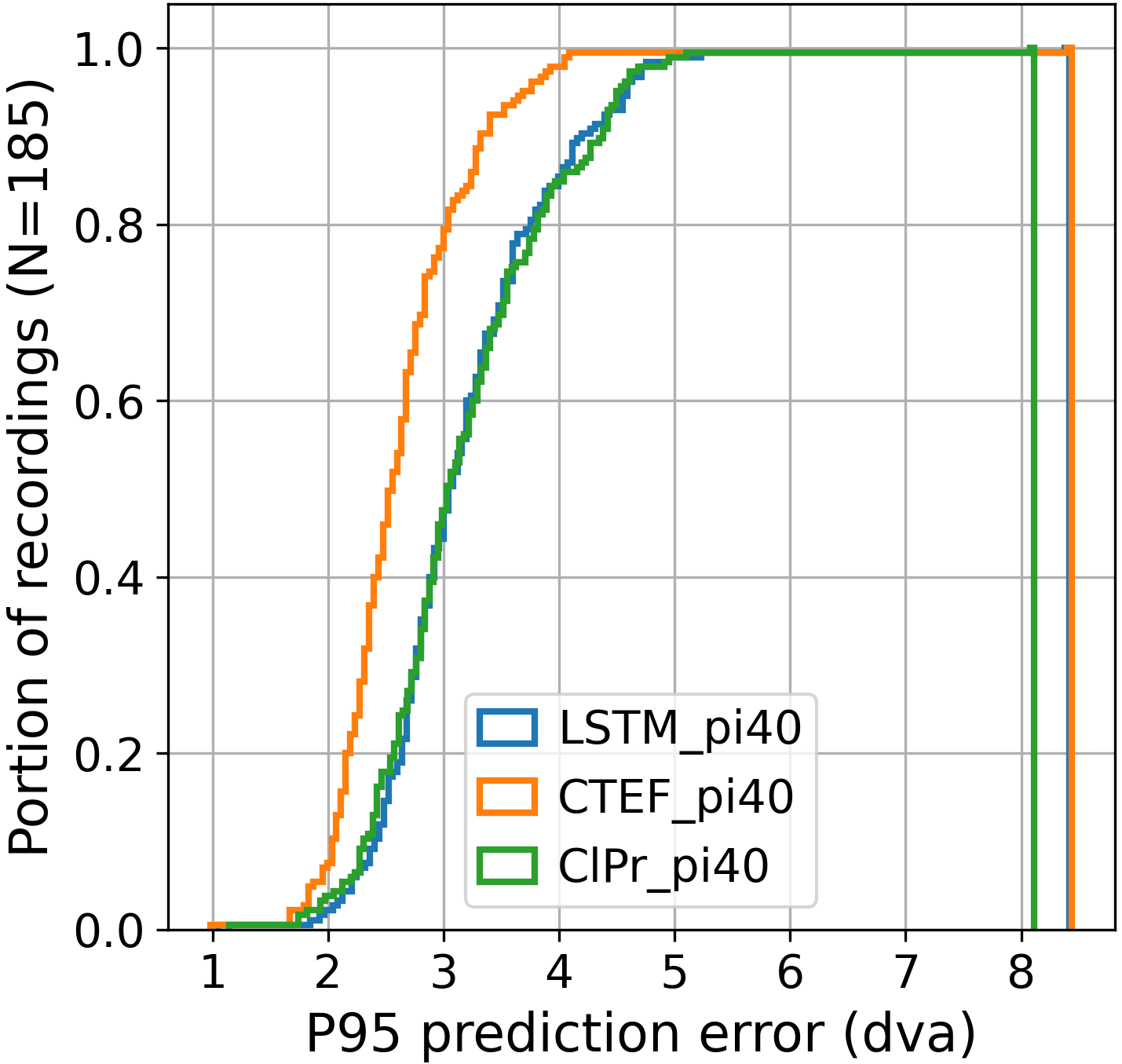}
            \captionsetup{justification=centering}
            \caption{\textbf{GBVR} \\ CDF of $P_{95}$ Event-Agnostic Errors}
            \label{fig:95_fullcdf_gbvr}
        \end{subfigure}
        \hfill
        \begin{subfigure}{0.32\textwidth}
            \centering
            \includegraphics[width=\textwidth]{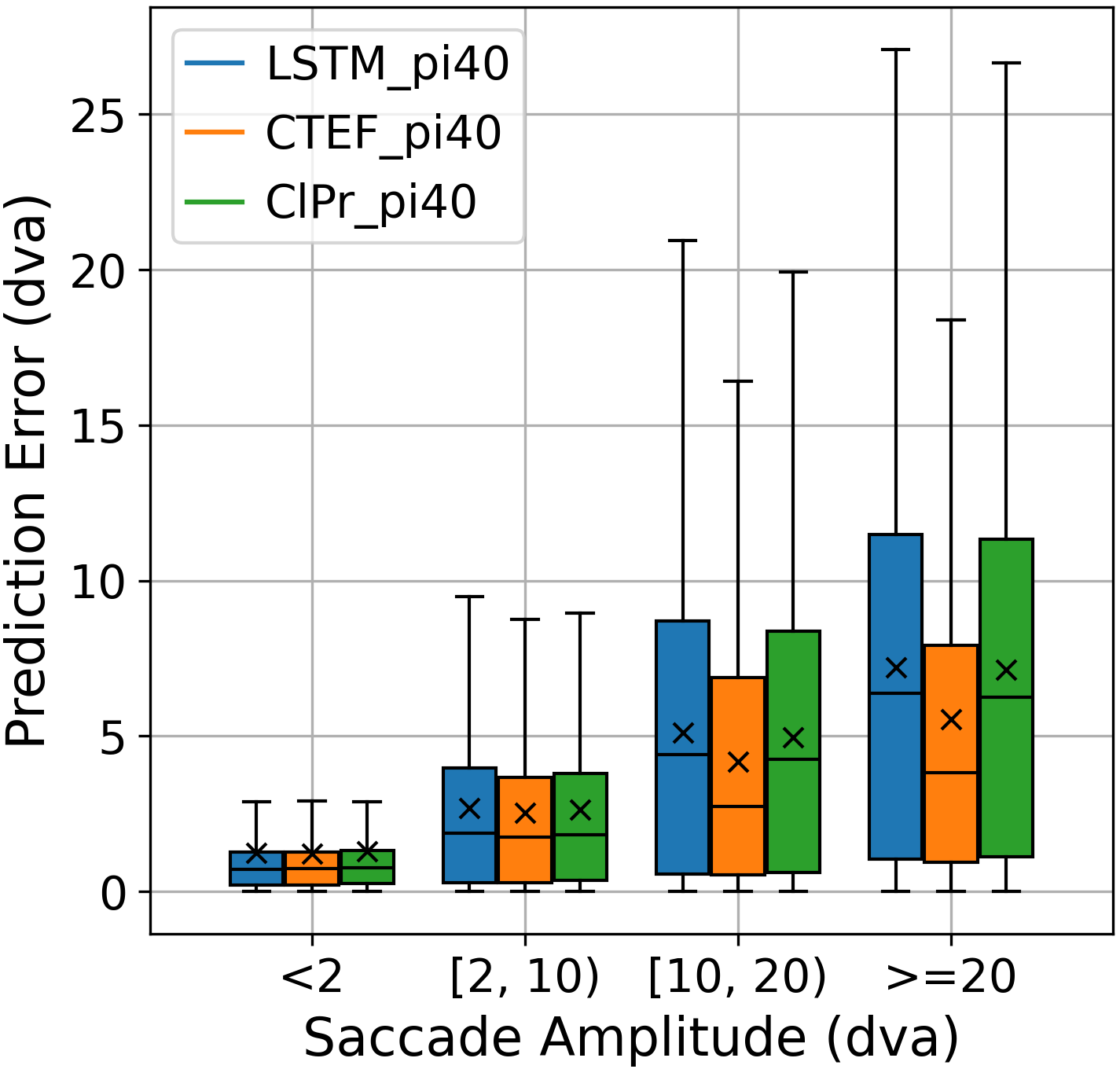}
            \captionsetup{justification=centering}
            \caption{\textbf{GBVR} \\ Saccade Prediction Errors}
            \label{fig:saccade_boxplot_gbvr}
        \end{subfigure}
        \hfill
        \begin{subfigure}{0.32\textwidth}
            \centering
            \includegraphics[width=\textwidth]{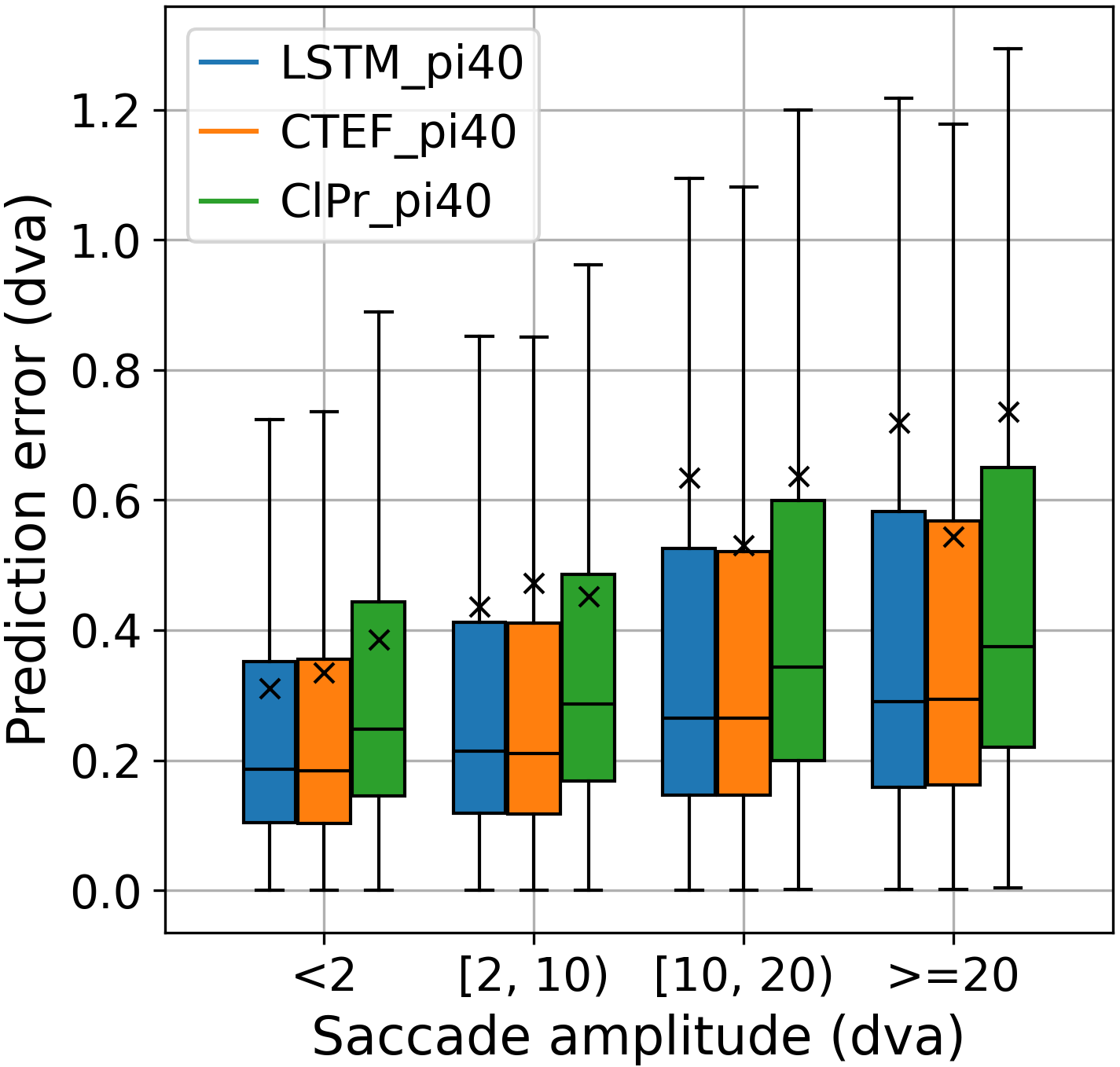}
            \captionsetup{justification=centering}
            \caption{\textbf{GBVR} \\ CEP Prediction Errors}
            \label{fig:cep_boxplot_gbvr}
        \end{subfigure}
        
    \caption{Comparative analysis of model prediction errors across Quest Pro (QP, top row) and GazeBase VR (GBVR, bottom row) datasets. (a, d) CDFs of $P_{95}$ event-agnostic errors. (b, e) Saccade prediction errors grouped by saccade amplitude. (c, f) Post-saccadic (CEP) error distributions categorized by the preceding saccade amplitude. In all boxplots, the horizontal line and cross symbol denote the median and mean, respectively.}
    \Description{Comparative analysis of model prediction errors across Quest Pro (QP, top row) and GazeBase VR (GBVR, bottom row) datasets. (a, d) CDFs of $P_{95}$ event-agnostic errors. (b, e) Saccade prediction errors grouped by saccade amplitude. (c, f) Post-saccadic (CEP) error distributions categorized by the preceding saccade amplitude. In all boxplots, the horizontal line and cross symbol denote the median and mean, respectively.}
    \label{fig:comparison_cdfs_boxplots}
\end{figure*}

We pooled out-of-sample test predictions across five folds, including 60 unique subjects for Meta Quest Pro (QP) and 185 for GazeBase VR (GBVR) to evaluate population-wide performance. Figures \ref{fig:95_fullcdf_qp} and \ref{fig:95_fullcdf_gbvr} present the cumulative distribution functions (cdf) of $P_{95}$ event-agnostic errors. The CTEF model showed the best performance, as its leftmost cdf indicates better consistency in managing worst-case errors across the subject population. Separate fixation plots are provided in the Supplemental Materials: since fixations represent the majority of data samples, their performance closely parallels the event-agnostic error distributions.

Prediction errors across saccades, grouped by amplitude, are shown in Figures \ref{fig:saccade_boxplot_qp} and \ref{fig:saccade_boxplot_gbvr}. All models performed similarly for small saccades ($<2^{\circ}$) and medium saccades ($[2^{\circ}-10^{\circ}]$). CTEF achieved the lowest errors for GazeBase VR across all amplitudes. LSTM showed a marginal (though statistically non-significant) advantage for larger saccades on the QP dataset. The post-saccadic error performance is illustrated in Figures \ref{fig:cep_boxplot_qp} and \ref{fig:cep_boxplot_gbvr}. There is a clear increase in the CEP error as the saccade amplitude increases for the GBVR dataset. In contrast, CEP errors remain largely invariant to the preceding saccade amplitude for the 90 Hz QP dataset. It suggests that at lower sampling frequencies, the post-saccadic error is independent of the preceding saccade amplitude. CTEF performs better for CEPs on the QP dataset, but the difference between LSTM and CTEF is minimal for GBVR.

\section{Gaze Prediction as a Function of Eye Movement Type Across Different PIs}
The 44 ms PI was prioritized for the main analysis as it offered the optimal balance between latency compensation and prediction accuracy. Results for 22 ms and 66 ms intervals are included in this section for comparison. We provide results from the first fold of the Quest Pro dataset as a representative example to demonstrate how the prediction error changes as PI increases. Table \ref{tab:comparison_QP} illustrates the performance differences among the three models. Although error growth is expected, the large amplitude saccade category exhibits the most significant sensitivity to increase in PI. For example, the LSTM $P_{50}$ error for large saccades increases from $1.96^{\circ}$ at $\text{PI}=22$ to $6.5^{\circ}$ at $\text{PI}=66$, nearly tripling. In contrast, the small saccade $P_{50}$ error remains stable, increasing by only $4\%$ (from $1^{\circ}$ 1 to $1.04^{\circ}$) across the same PI values. Given that small saccades are among the most challenging eye movements to predict, many architectures fail to capture them effectively. CTEF and LSTM both outperform ClPr. Minor fluctuations in the performance gap occasionally occur. The level of significance for other PIs should be assessed separately to confirm these trends. 

\begin{table*}[h]
    \centering
    \caption{Comparative analysis of gaze prediction errors across three prediction intervals ($\text{PI} \in \{20,40,60\}$ ms). Results are evaluated on the Quest Pro dataset (Fold 1). Bold values (highlighted in green) indicate the lowest prediction error for each eye movement category.}
    \begin{tabular}{|l|l|c|c|c|c|c|c|c|c|}
        \hline
        % Header setup
        \multirow{2}{*}{\textbf{PI}} & \multirow{2}{*}{\textbf{Model}} & \multicolumn{2}{|c|}{\textbf{Fixation}} & \multicolumn{2}{c|}{\textbf{Large Saccade}} & \multicolumn{2}{c|}{\textbf{Small Saccade}} & \multicolumn{2}{c|}{\textbf{CEP}} \\
        \cline{3-10} % Use \cline instead of \cmidrule(lr)
        & & $P_{50}$ & $P_{95}$ & $P_{50}$ & $P_{95}$ & $P_{50}$ & $P_{95}$ & $P_{50}$ & $P_{95}$ \\
        \hline

        % Data rows
        \multirow{3}{*}{\textbf{22}} & LSTM & 0.4 & 0.95 & \textbf{\textcolor{mygreen}{1.96}} & \textbf{\textcolor{mygreen}{5.03}} & \textbf{\textcolor{mygreen}{1}} & \textbf{\textcolor{mygreen}{1.55}} & 0.48 & 1.3 \\
        \cline{2-10} % Add a \cline here to separate the multirow section visually
        & CTEF & \textbf{\textcolor{mygreen}{0.4}} & \textbf{\textcolor{mygreen}{0.9}} & 1.98 & 5.25 & 1 & 1.56 & \textbf{\textcolor{mygreen}{0.47}} & \textbf{\textcolor{mygreen}{1.15}} \\
        \cline{2-10}
        & ClPr & 0.46 & 1.05 & 2 & 5.05 & 1.08 & 1.56 & 0.52 & 1.3 \\
        \hline \hline
        
        % Data rows
        \multirow{3}{*}{\textbf{44}} & LSTM & 0.45 & 1.47 & 4.07 & 10.23 & \textbf{\textcolor{mygreen}{1.02}} & \textbf{\textcolor{mygreen}{1.67}} & 0.65 & 3.16 \\
        \cline{2-10} % Add a \cline here to separate the multirow section visually
        & CTEF & \textbf{\textcolor{mygreen}{0.45}} & \textbf{\textcolor{mygreen}{1.4}} & \textbf{\textcolor{mygreen}{4.03}} & \textbf{\textcolor{mygreen}{10.20}} & 1.05 & 1.73 & \textbf{\textcolor{mygreen}{0.65}} & \textbf{\textcolor{mygreen}{3.06}} \\
        \cline{2-10}
        & ClPr & 0.54 & 1.74 & 4.1 & 10.07 & 1.12 & 1.82 & 0.73 & 3.62 \\
        \hline \hline

        % Data rows
        \multirow{3}{*}{\textbf{66}} & LSTM & 0.53 & 2.46 & \textbf{\textcolor{mygreen}{6.5}} & \textbf{\textcolor{mygreen}{13.61}} & \textbf{\textcolor{mygreen}{1.04}} & \textbf{\textcolor{mygreen}{1.62}} & 1.1 & 5.7 \\
        \cline{2-10} % Add a \cline here to separate the multirow section visually
        & CTEF & \textbf{\textcolor{mygreen}{0.52}} & \textbf{\textcolor{mygreen}{2.41}} & 6.55 & 13.67 & 1.12 & 1.71 & \textbf{\textcolor{mygreen}{1.07}} & \textbf{\textcolor{mygreen}{5.6}} \\
        \cline{2-10}
        & ClPr & 0.7 & 2.63 & 6.71 & 13.55 & 1.14 & 1.92 & 1.49 & 5.7 \\
        
        \hline % Final horizontal line
    \end{tabular}
    \label{tab:comparison_QP}
\end{table*}

\section{Visualizations: Gaze Prediction as a Function of Individual Differences}
\begin{figure*}
    \centering
    \begin{subfigure}{0.40\textwidth}
        \includegraphics[width=\textwidth, height=4.25cm]{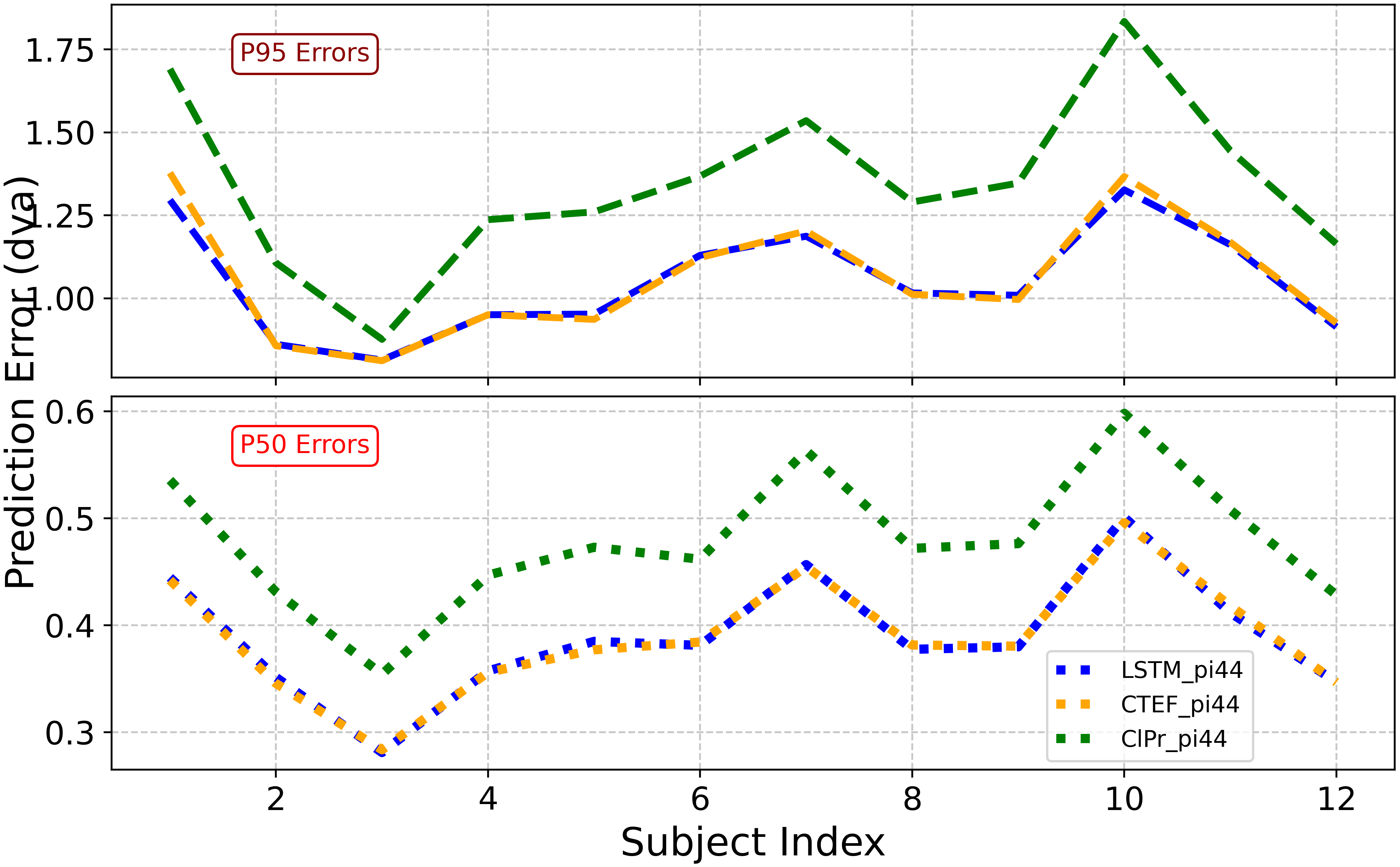}
        \caption{Per-Subject Fixation Error Profiles}
        \label{fig:pq_fix_sub}
    \end{subfigure}
    \begin{subfigure}{0.40\textwidth}
        \includegraphics[width=\textwidth, height=4.25cm]{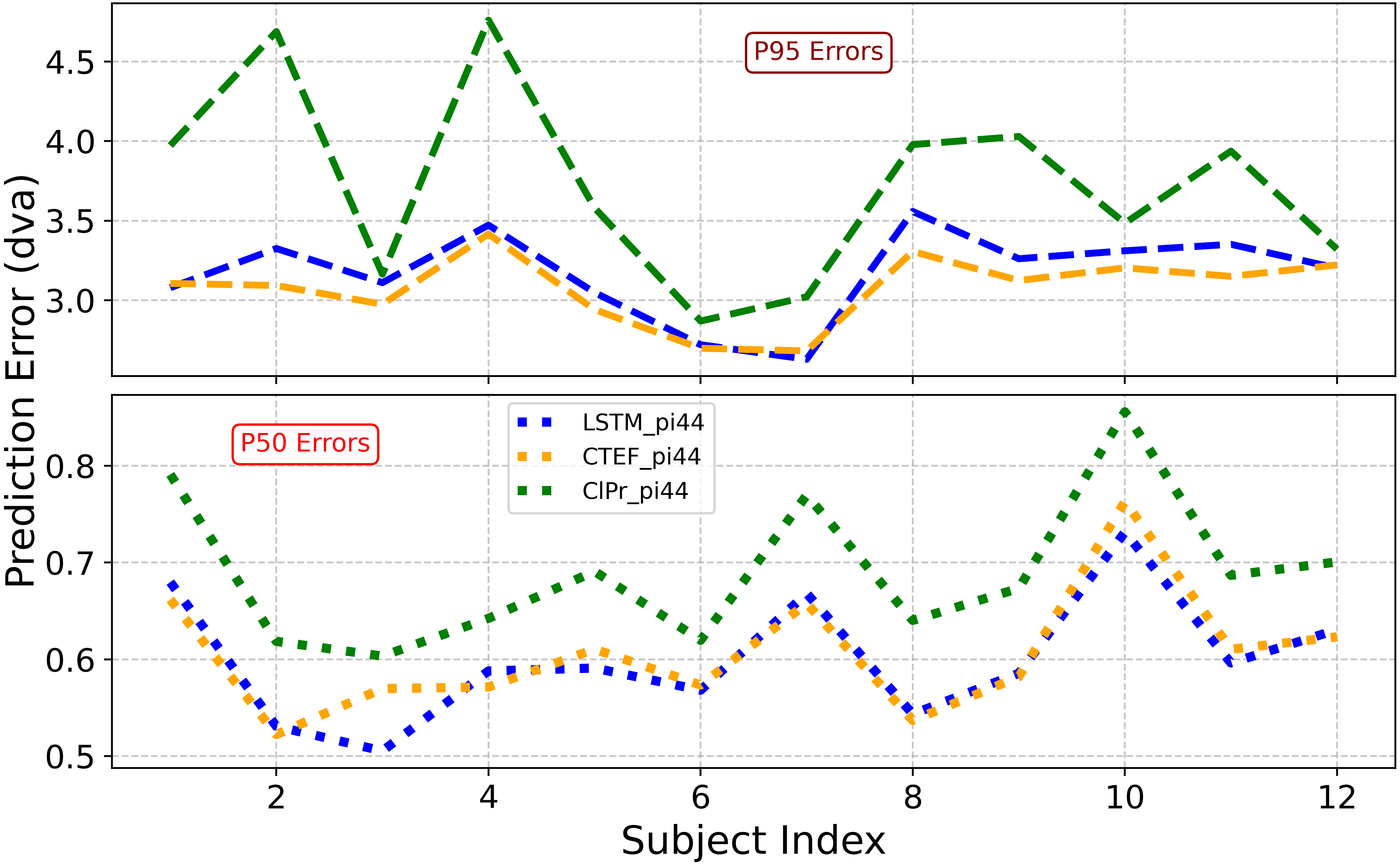}
        \caption{Per-Subject CEP Error Profiles}
        \label{fig:pq_cep_sub}
    \end{subfigure}

    \begin{subfigure}{0.40\textwidth}
        \includegraphics[width=\textwidth, height=4.25cm]{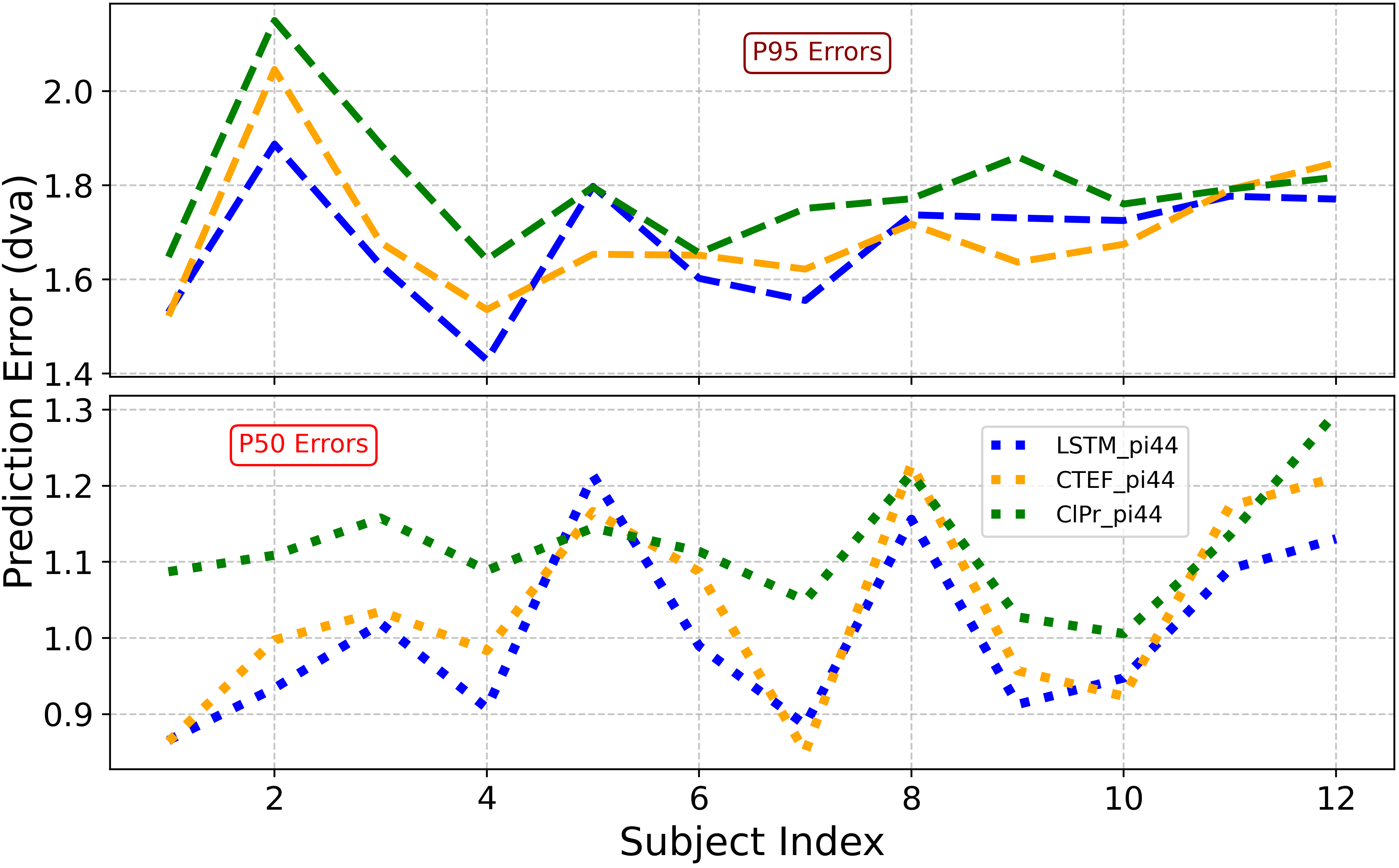}
        \caption{Per-Subject Small Saccade Error Profiles}
        \label{fig:pq_sm_sac_sub}
    \end{subfigure}
    \begin{subfigure}{0.40\textwidth}
        \includegraphics[width=\textwidth, height=4.25cm]{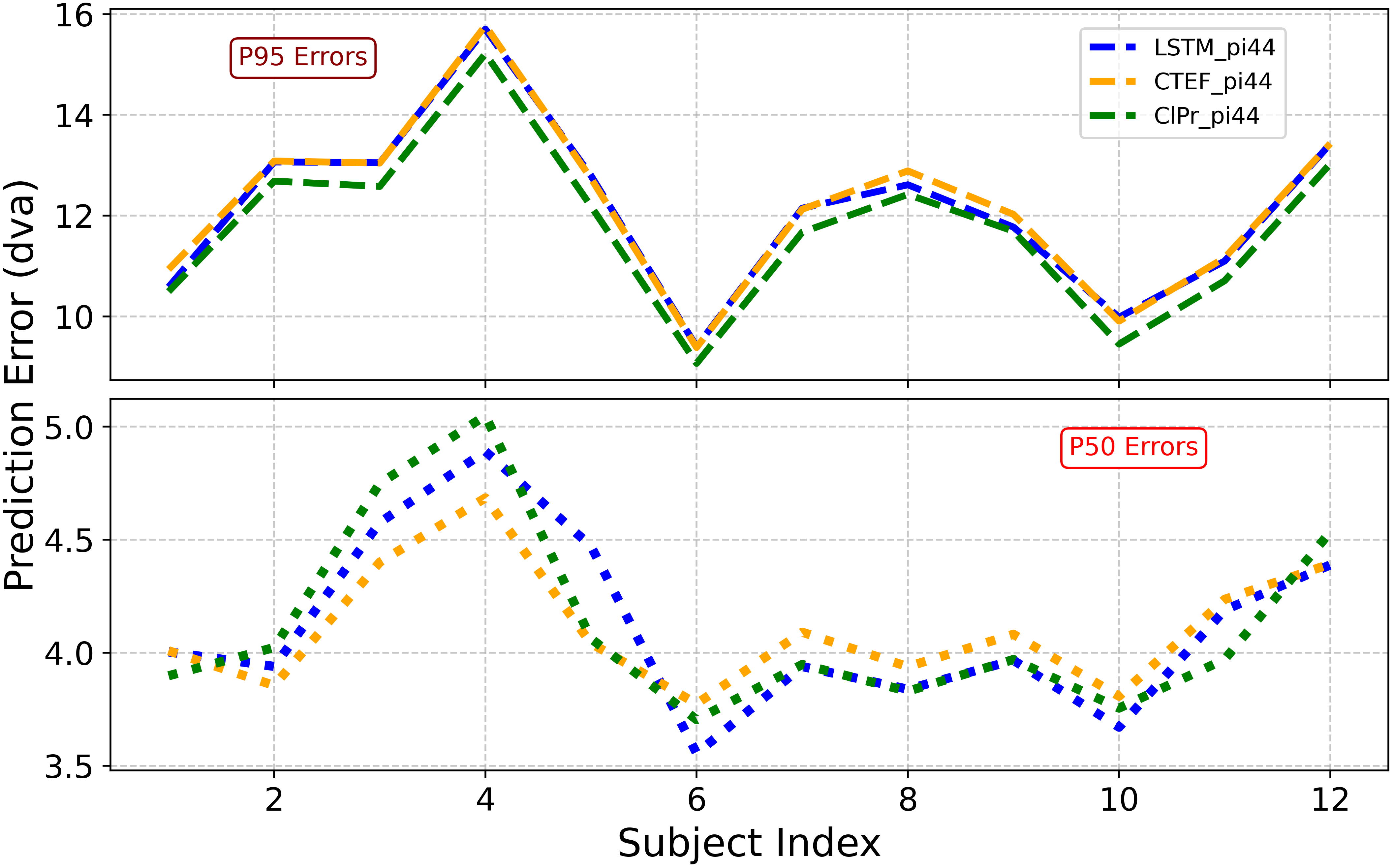}
        \caption{Per-Subject Large Saccade Error Profiles}
        \label{fig:pq_lr_sac_sub}
    \end{subfigure}
    \caption{Subject-wise gaze prediction error profiles ($\text{PI}=44$ ms) on the Quest Pro dataset (Fold 1). Error distributions are shown for the 95th ($P_{95}$, top) and 50th ($P_{50}$, bottom) percentiles across 12 subjects (24 recordings total). Panels categorize errors by eye-movement type: (a) Fixations, (b) Critical Evaluation Periods (CEPs), (c) Small Saccades, and (d) Large Saccades. The profiles highlight the degree of inter-subject variability across eye-movements.}
    \Description{Subject-wise gaze prediction error profiles ($\text{PI}=44$ ms) on the Quest Pro dataset (Fold 1). Error distributions are shown for the 95th ($P_{95}$, top) and 50th ($P_{50}$, bottom) percentiles across 12 subjects (24 recordings total). Panels categorize errors by eye-movement type: (a) Fixations, (b) Critical Evaluation Periods (CEPs), (c) Small Saccades, and (d) Large Saccades. The profiles highlight the degree of inter-subject variability across eye-movements.}
    \label{fig:all_pq_profiles}
\end{figure*}

Continuing with the first fold of the Quest Pro dataset, subject error profiles were generated for fixations, CEPs, small and large saccades (Fig. \ref{fig:all_pq_profiles}). These profiles reveal significant individual differences in performance. For fixations, the $P_{50}$ and $P_{95}$ profiles are highly consistent across subjects. This suggests that predictability is an inherent property of the individual subject (see Fig.\ref{fig:pq_fix_sub}). The CEPs lacked this consistency. Subjects with high $P_{50}$ errors did not necessarily demonstrate the highest extreme-case $P_{95}$ errors. For this metric, LSTM and CTEF showed better $P_{50}$ performance. CTEF remained the most accurate based on $P_{95}$ analysis (Fig. \ref{fig:pq_cep_sub}). 

For small saccades, the $P_{50}$ profile patterns vary markedly across both subjects and models (Fig. \ref{fig:pq_sm_sac_sub}). This variation suggests that the models focus on different features when making predictions. For large saccades (Fig. \ref{fig:pq_lr_sac_sub}), CTEF demonstrated the best $P_{50}$ performance, although the $P_{95}$ trends remained nearly identical across all three models. These subject-specific trends reveal performance nuances lost in aggregate data, providing an essential diagnostic layer for benchmarking eye movement prediction.

\section{Computational Cost Analysis}
Our evaluations were conducted on an NVIDIA RTX A4000 GPU (CUDA 12.2) to quantify the relative performance scaling. While we acknowledge that absolute metrics (e.g., raw latency) will differ on mobile VR hardware, comparative performance trends observed on high-end silicon typically provide a reliable proxy for mobile architectures. Table \ref{tab:model_times} presents the resulting inference speeds.

\begin{table}[ht]
    \centering
    \caption{ Inference performance of the evaluated gaze prediction models. Mean ($\mu$) and standard deviation ($\sigma$) of inference times are measured over 100 consecutive runs. Input window sizes are reported as (channels, timesteps), and the total number of model parameters is given in millions (M).}
    \begin{tabular}{|l|c|c|c|c|}
        \hline % Top horizontal line
        \multirow{2}{*}{\textbf{Model}}& \multirow{2}{*}{\textbf{Input Size}} & \multicolumn{2}{c|}{\textbf{Inference Time}} & \multirow{2}{*}{\textbf{Parameters}} \\
        \cline{3-4}
        & & $\mu$ & $\sigma$ & \\
        \hline % Line separating header from data
        LSTM & (5, 12) & 0.55 ms & 0.005 & 4.02 M \\
        \hline
        CTEF & (4, 6)  & 0.97 ms & 0.01 & 2.02 M \\
        \hline
        ClPr & (4, 6)  & 0.54 ms & 0.01 & 0.745 M \\
        \hline
    \end{tabular}
    \label{tab:model_times}
\end{table}

Each model performed 10 warm-up forward passes before timing began. The mean ($\mu$) and standard deviation ($\sigma$) of the inference time were then calculated over 100 independent iterations. The costs of pre-processing were excluded from these measurements (see Table \ref{tab:model_times}). In contrast, the inference time of OPKF is estimated at approximately 0.44 ms ($\sigma=0.002$) for a 40 ms PI on 1000 Hz data, based on the parameters reported in study by  \citet{melnyk_gp}. This illustrates the significant advantage that mathematical models maintain over DL models in terms of both speed and computational footprint. The OPKF's computational cost will decrease at 90 Hz because it processes approximately 10 times fewer samples (4 vs. 40) to predict the same $\text{PI}=40$ ms, as the update interval relaxes from 1 ms to 11.1 ms.

Beyond raw speed, the limited memory bandwidth of Qualcomm Systems-on-Chip (SoC), such as the Snapdragon XR2+ Gen 1 powering the Meta Quest Pro headset, is a major bottleneck. This constraint is particularly restrictive for transformer-based models, whereas LSTMs and mathematical models like OPKF maintain a much smaller memory footprint. The CTEF architecture is lightweight, but its self-attention mechanisms still require optimizations such as quantization or pruning to run smoothly on a headset.
%% End of the file.
\end{document}